\documentclass[aps,pra,twoside,twocolumn,10pt]{revtex4-1}
\usepackage[colorlinks=true, citecolor=blue, urlcolor=blue ]{hyperref}
\usepackage{epsfig,newlfont,amssymb,amsfonts,amsmath,bm,subfigure,palatino,mathtools,amsthm,braket,soul,enumitem,color,graphics,graphicx,times,physics}
\usepackage[normalem]{ulem}

\begin{document}
\title{Refrigeration via purification through repeated measurements}

\author{Tanoy Kanti Konar, Srijon Ghosh, Aditi Sen(De)}
\affiliation{Harish-Chandra Research Institute,  A CI of Homi Bhabha National
Institute, Chhatnag Road, Jhunsi, Prayagraj - 211019, India}

\begin{abstract}

We design a measurement-based quantum refrigerator with an arbitrary number of qubits situated in a
one-dimensional array that interact through variable-range \(XY\) interactions. The method proposed is based on repeated evolution followed by a measurement on the single accessible qubit, which has the potential to reduce the temperature in the rest of the subsystems, thereby demonstrating cooling in the device. The performance of the refrigerator is quantified by the fidelity of each local subsystem with the ground state of the local Hamiltonian and the corresponding probability of success. We identify system parameters, which include the interaction strength, range of interactions, initial temperature of each qubit, and the position of the measured qubit, so that the fidelities of all the unmeasured qubits approach unity with a nonvanishing probability. We observe that although strong interactions during evolution are required to achieve cooling, the long-range interactions typically deteriorate the performance of the refrigerator, which indicates that interactions are not ubiquitous. We report the scalability and the saturation property of the success probability with respect to the system size, which turns out to be independent of the involved system parameters and the number of repeated measurements. Furthermore, we show that the number of subsystems which can be cooled changes depending on the odd or even number of sites in the refrigerator. We argue that the distribution of entanglement between unmeasured qubits can give a possible explanation of the dependence of cooling process on the measured and unmeasured sites.

\end{abstract}

\maketitle




\section{Introduction}

In the last few years, the breakneck modernization of technology has created several essential and pertinent directions in the field of quantum thermodynamics \cite{gemmer2004, binder2018, lekscha2018, deffner2019}. Rapid developments also demand designing of quantum devices which perform more efficiently than the existing ones and can be built by using currently available technologies. In all the discovered thermal machines, like quantum battery \cite{Alicki}, quantum transistor \cite{karl2016}, diode \cite{ordonez2017} and  quantum refrigerator \cite{linden2010, allahverdyan2010, levy2012,  hewgill2020}, significant enhancements can be shown to be achieved in the quantum domain compared to classically available devices. Most importantly, these machines designed by the Dicke system \cite{PoliniPRL}, low as well as higher dimensional systems \cite{correa2014,wang2015,silva2016,konar2022}, mechanical resonators \cite{li2011} and quantum spin models \cite{hewgill2020,konar2021} have successfully been engineered in physically realizable substrates like cold atoms \cite{duan2003}, nuclear magnetic resonances \cite{kose2019}, trapped ions \cite{maslennikov2019}, atom-cavity systems \cite{mitchison2016} and the superconducting circuits \cite{manikandan2019}. In spite of all these advancements, it has also been realized that there are several avenues  by which the efficiencies of these devices can be improved. 

Three interacting quantum systems \cite{henrich2007, linden2010,huber2015,das2019,konar2021}, each connected to a non-interacting Markovian thermal bath having different temperatures can be a possible set-up of a small quantum refrigerator. Typically, for lowering the temperature of the \emph{target system}, the dynamics of the system is controlled in such a manner that thermal energy is transferred to the hot bath from a cold one with the help of another  heat bath, known as \emph{work reservoir} \cite{Petruccione}, both in the steady state as well as in the transient regimes. Precisely, the main aim towards such a functioning of a refrigerator is to keep the state of the target system in the available ground state, so that the local temperature of the system becomes minimum, thereby achieving the cooling. Numerous works have recently been carried out in this direction concentrating mainly on the system consisting of qubits, qubit-qutrit \cite{tonner2005, linden2010, bhandari2021} and also higher dimensional spin systems \cite{konar2022}. In addition, these steady state cooling procedures can be accomplished either by external energy sources or by disallowing sources of external control, thereby obtaining self-contained heat engines \cite{linden2010, correa2013} or in a periodically driven system \cite{niedenzu2018} or via repeated collision \cite{dechiara2018} or with reverse-coupling mechanism \cite{silva2015}. 

All the aforementioned designing of quantum refrigerators is confined to the fact that the state of the target system can attain the corresponding ground state from an arbitrary thermal state, thereby arriving at a lower temperature from its initial one by means of open-system evolution described by the Gorini–Kossakowski–Sudarshan–Lindblad (GKSL) master equation \cite{Petruccione, rivas2012}. In this work, we choose a completely distinct route -- starting from local spin Hamiltonian, a phenomena of cooling i.e., a refrigeration of several quantum systems are achieved via repeated unitary evolution of the entire system governed by quantum prototypical spin models followed by a measurement on a single accessible qubit. From a different perspective, the repeated measurement scenario was first considered in the context of purifying a subsystem from the mixed state by performing frequent measurements on another subsystem \cite{nakazato2003, nakazato2004}. Almost at the same time,  an unitary evolution-based cooling protocol of a nearest-neighbor spin system has been proposed \cite{burgarth2007,burgarth2008} in which 
by using a set of non-interacting qubits (controller system),   the sequential interactions between  one of the controllers and the spin model leads to cooling by applying
an unitary swap operator. It was later on used to cool a mechanical resonator by performing measurements on the flux qubit which are in contact with the resonator \cite{li2011, pyshkin2016, montenegro2018, puebla2020} and also in the spin systems \cite{zhang2019}. Recent work generalizes the situation for double resonators, thereby achieving simultaneous cooling \cite{yan2021, yan2022}. It was also shown that in this process of repeated measurements, bipartite as well as multipartite entangled states can be generated \cite{nakazato2003, nakazato2004, wu2004, qiu2012}. 


In this proposal, the state of the measured qubit is initially taken to be the ground state of the local Hamiltonian, while unmeasured qubits are prepared in the thermal state of the same local Hamiltonians. At a later time, the interaction between measured as well as unmeasured qubits are turned on according to the anisotropic variable range $XY$  Hamiltonian  up to a certain interval of time to generate the correlation between the systems. We now address the following question: \\
\emph{Is it possible to refrigerate every individual unmeasured qubit of the system by measuring only a particular subsystem repeatedly?} \\
We answer it affirmatively.  We report that by choosing the system parameters, time of evolution and the measurement basis appropriately, it is possible to decrease the temperature of every individual qubit with a finite probability when the dynamics is governed by both short- and long-range interactions. In particular, the successful cooling process is measured by computing the fidelity between the time evolved reduced subsystem after measurements and its corresponding ground state of each local subsystem. Notice that the procedure of cooling is different from the quantum Zeno effect \cite{misra1977}, which deals with frequent measurement, thereby slowing down the dynamics of the system. 

When a refrigerator consists of three spin-$\frac{1}{2}$ particles, we derive analytical forms of fidelity after arbitrary rounds of evolution and measurements. Numerical simulations reveal that when an one-dimensional (1D) array of spins composed of at most eight spin-$\frac{1}{2}$ particles represents a refrigerator, the fidelity depends crucially on the range of interactions, system parameters and the position of the measured qubits in the array. Although the protocol is probabilistic, 
we find that the probability of successfully achieving cooling saturates  to a finite nonvanishing value with the number of measurements and we also report its scaling with the system size, 
when cooling occurs for all the unmeasured qubits which are independent of the system parameters.

Some of the previous works \cite{nakazato2003, nakazato2004} have shown that one can  generate a multiparty entangled state by measuring on a single qubit repeatedly. We show here that the cooling procedure can be connected to the distribution of entanglement \cite{coffman2000, Dhar2017} among unmeasured qubits. In particular, when the sum of entanglement between the unmeasured qubits decreases, it immediately implies that the refrigerator works. This is due to the fact that quantum correlation created at the initial steps induces cooling in the system although the procedure becomes successful only when the final state  produced between unmeasured qubits is fully separable state.

The paper is organised as follows: In Sec. \ref{sec:model}, we introduce the model of refrigeration for arbitrary number of qubits interacting through the variable range interaction and define the figure of merits. In the next section (Sec. \ref{sec:ana}), we derive a compact analytical form of fidelity for a three qubit refrigerator and compare the performance depending on the position of the measured qubits. Proceeding further in Sec. \ref{sec:refrimorequbits}, we examine the effects of measurements and coupling strength when the initial states of the subsystems are maximally mixed for a refrigerator built up with more than three spin-$\frac{1}{2}$ particles. Sec. \ref{sec:initial_temp_dependence} reports the efficiencies of the refrigerator when the unmeasured qubits are prepared in the thermal state with moderate temperature. In Sec. \ref{sec:spatial}, we present the consequences of measurements performed on different spatial positions on cooling and finally conclude in Sec. \ref{sec:conclusion}. 

  \begin{figure}
     \centering
     \includegraphics[height=4 cm,width=8.8 cm]{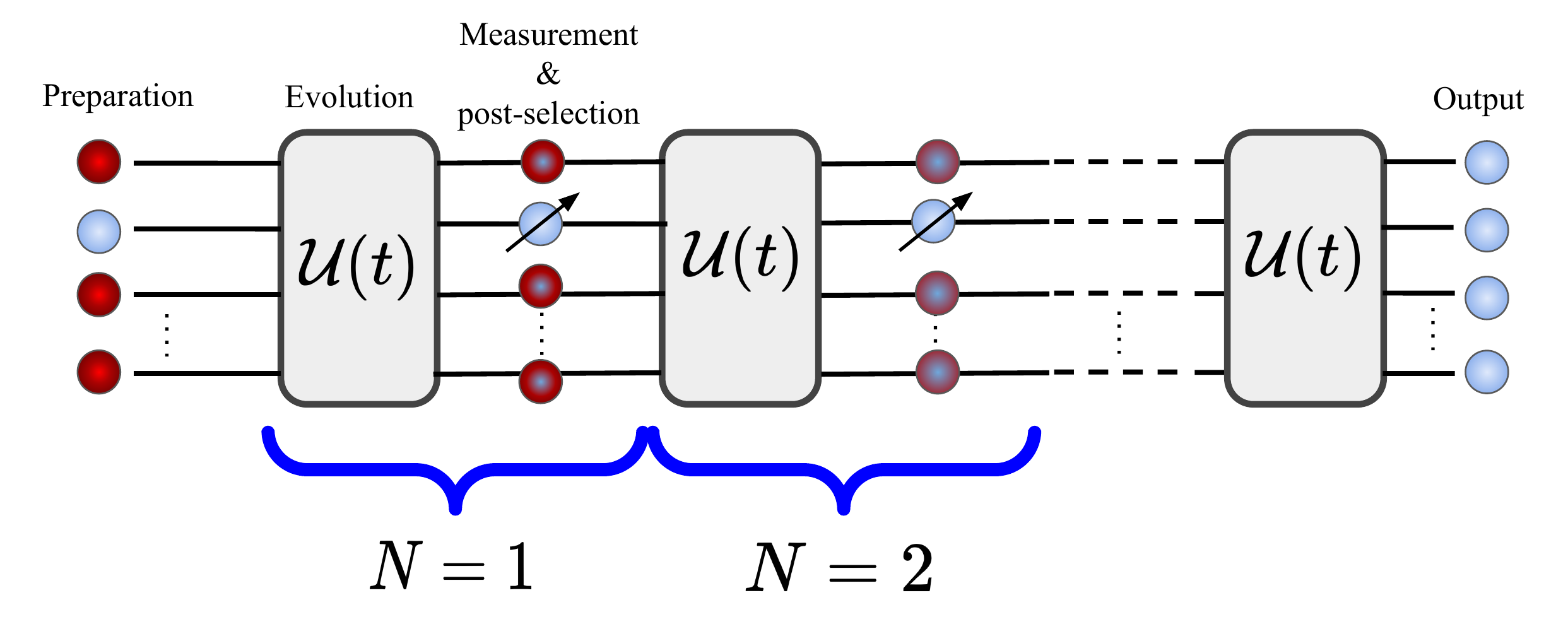}
     \caption{(Color online) Steps of a measurement-based refrigerator. 
     An arbitrary qubit (blue)  which is initially prepared as the ground state of the local Hamiltonian  is chosen for the measurement while the rest of qubits  (red) are initially in the thermal state of a local Hamiltonian with finite temperature. The system evolves according to the unitary operator $\mathcal{U}(t)$ followed by a measurement  on a single qubit which is in the ground state. By properly choosing the desired output state of the measured qubit, we again let the whole system evolve according to the same unitary.  The aim of such repeated evolution followed by measurements is to project all (some) of qubits to the ground state of the local Hamiltonian, thereby achieving the cooling. We show that it is indeed possible to achieve.}
     \label{fig:figA}
 \end{figure}

\section{Repeated Measurement-based  Refrigerator model}
\label{sec:model}

\begin{figure}
     \centering
     \includegraphics[height=4 cm,width=8.2 cm]{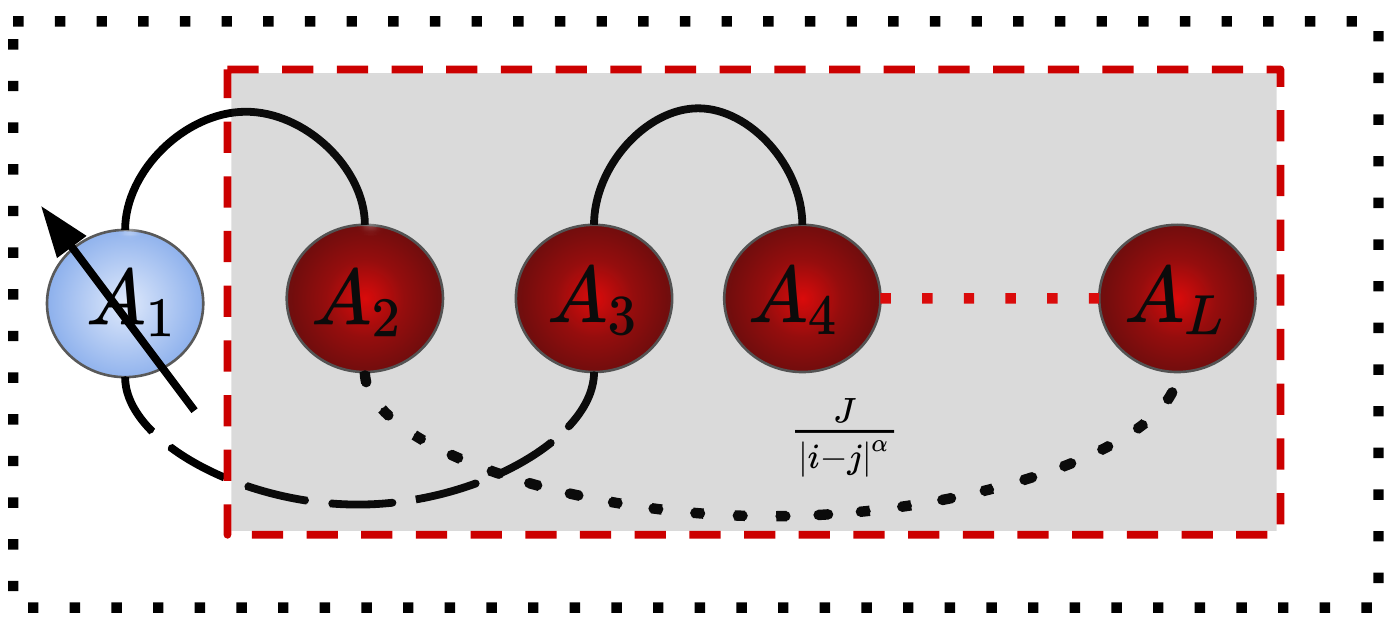}
     \caption{(Color online) Schematic diagram of the spin-chain where during dynamics, the system evolves according to the \(H_{ev}\), containing variable range interacting Hamiltonian, \(H_{int}\) which scales as \(\frac{J}{|i-j|^\alpha}\), \(i\) and \(j\) being the sites of the spins. An accessible qubit is chosen for the measurement and is initially prepared as the ground state of the local Hamiltonian while the rest of the inaccessible qubits (shaded region) are initially in the thermal state with finite temperature.} 
     \label{fig:fig1}
 \end{figure}

Let us first introduce the set-up of a refrigerator based on repetitive quantum measurements (Fig. \ref{fig:figA}). The model consists of \(L\) spin-$\frac{1}{2}$ particles, denoted as \(A_1\), \(A_2, \ldots A_L\), governed by the local Hamiltonian, \(H_{A_{i}}\) (\(i= 1, 2, \ldots, L\)) in an one-dimensional  array depicted in Fig. \ref{fig:fig1}. The measurement is performed on a single accessible qubit placed in any arbitrary position of the array, referred to as \(A^M_j\) (\(j \ne i\)), which is prepared in the ground state $|\mathcal{G}_{A^{M}_j} \rangle$ of its local Hamiltonian \(H_{A^{M}_j}\). The rest of the unmeasured qubits are prepared in the thermal states of there respective local Hamiltonians $H_{A_{i}}$, $(i = 1,2 \ldots L,\,\, i\neq j)$. Hence the initial state of the total $L$-party system reads as
\begin{equation}
    \rho(0) =  |\mathcal{G}_{A^{M}_j} \rangle \langle \mathcal{G}_{A^{M}_j}|  \otimes \rho_{in}^{L-1}.
    \label{initial_state_general}
\end{equation}
where $\rho_{in}=\bigotimes_{i\neq j, i=1}^{L} \rho_{A_i}$ is the initial state of the unmeasured qubits. Notice that the initial state is fully separable, having vanishing multiparticle as well as bipartite entanglement \cite{horodecki2009, vidal2002,plenio2005}. When $t>0$, the system evolves according to the Hamiltonian, \(H_{ev}\), consisting of  \(\sum_{i} H_{A_{i}} + H_{A^{M}_j} + H_{int}\) for a certain interval of time $t$, where $H_{int}$ represents the interaction term between $L$ spin-$\frac{1}{2}$ particles (see Fig. \ref{fig:fig1}). In this situation, it is reasonable to expect that the corresponding global unitary operator, $\mathcal{U}(t) = \exp(- i H_{ev} t)$, can be responsible to create quantum correlation between $L$ subsystems. The evolution is followed by an arbitrary projective measurement at site \(A^M_j\) with the set of projection operators \(\{\Pi^M_k\}_{k=1}^2\), whose elements are the ground and the excited states of \(H_{A^M_j}\). A post-selection of the output state is performed to ensure that the measured qubit, \(A^M_j\) is projected to the ground state i.e., $\Pi^M_1=|\mathcal{G}_{A^{M}_j} \rangle \langle \mathcal{G}_{A^{M}_j} |$ while we discard the process if the excited state clicks and start the protocol from the beginning. In case of the desired outcome, \(\Pi_1^M\), the entire process is repeated for arbitrary number of times, say \(N\) times, until the individual \(L-1\) qubit reach their respective ground states of the initial Hamiltonian, \(|\mathcal{G}_{A_{i}}\rangle\). Starting from the thermal state with an arbitrary high temperature, the above process demonstrates that the repeated evolution for a short period of time and projective measurements can drive all (some of) the unmeasured subsystems to reach the ground states, thereby lowering the temperature (energy) of the individual subsystem as well as the entire system. We call this procedure as refrigeration. In other words, cooling occurs successfully when purification of thermal states takes place, thereby reaching to pure states having minimum energy. 
Note that  in contrast to the well-known Zeno-measurement, we consider the time of the dynamics $t$ to be nonvanishing and finite as well as the repetition of measurements, $N$ to be large but finite.  
As argued in Ref. \cite{nakazato2003} a purification of the unmeasured part of the system occurs due to the projection into the eigenstate corresponding to the maximum eigenvalue of the non-unitary evolution operator. 
With increasing number of measurements, such eigenstate dominates which depends on the choice of parameter and  hence  we can project unmeasured state to a different purified state by tuning system parameters suitably. This picture is similar to the above proposed protocol where change of spatial position of measurement  can lead to different eigenstates of the non-unitary evolution, thereby leading to  distinct fidelities of the unmeasured qubits. We will also show that the successful implementation of the cooling process  can also be explained via entanglement properties of the unmeasured qubits. 

After the completion of the whole process of repeated evolution and measurements for a particular interval of time, $t$, the resultant state in case of desired outcome from the measurement can be represented as
\begin{eqnarray}
\rho(t) = \frac{1}{p_N}(\Pi^M_1 \mathcal{U})^{N} \rho(0) (\Pi^M_1 \mathcal{U}^{\dagger})^{N}.
\end{eqnarray}
Here \(p_N\) is the probability of a successful implementation of cooling at each step  which can be calculated  as 
\begin{eqnarray}
p_N=\Tr [(\Pi^M_1 \mathcal{U})^{N} \rho(0) (\Pi^M_1 \mathcal{U}^{\dagger})^{N}],
\end{eqnarray}
 when the measurement outcome at each step is \(\Pi_1^M\) at site \(A^M_j\). 
We calculate the reduced state of each unmeasured subsystem by tracing out the other $L-1$ qubits, given by
\begin{equation}
    \rho_{A_i}(t) = \Tr_{\bar{A_i}}[\rho(t)],
    \label{reduced_state}
\end{equation}
 where $\overline{A_i}$ represents all the qubits except \(A_i\).
 As mentioned before, the main focus in building a refrigerator is to arrive at the ground state of as many unmeasured qubits as possible. To measure the performance of the refrigeration, we compute the fidelity of each qubit after an arbitrary step, $N$, in the ground state of the local Hamiltonian, $H_{A_i}$, which  mathematically reads as
\begin{equation}
    \mathcal{F} \equiv \mathcal{F}_{A_{i}}(N) = \langle \mathcal{G}_{A_{i}}| \rho_{A_i}(t) |\mathcal{G}_{A_{i}} \rangle.
\end{equation}
where $|\mathcal{G}_{A_{i}}\rangle$ is the ground state of the qubit, $A_i$, corresponding to the initial Hamiltonian $H_{A_{i}}$. 
Note that \(\mathcal{F}\) also depends on the tuning parameter of the system which can be functions of different parameters involved in the system like the interaction strength, range of interactions, the strength of the magnetic field,  initial temperature of the unmeasured qubits. Since our aim is to build a proper refrigerator, we continue the process till the fidelities for all (some) of the qubits approach to unity. 
Hence, during investigations, we have to choose  suitable (optimized) system parameters as well as projection operators so that the cooling is achieved.
Moreover, each step consists of measurements and the performance of the refrigerator 
can only be ensured if one obtains a nonvanishing probability, \(p_N\) after $N$ steps for some subsystems.


\subsection{ Realizing refrigeration with quantum spin models}

The refrigerator discussed above has three important components -- (1) initial state of the individual qubit governed by a local Hamiltonian; (2) the evolution operator especially, the interaction part, which is responsible for nontrivial  dynamics; (3) the projective measurement. Let us illustrate each ingredient of a refrigerator for a specific implementation considered in this manuscript.

{\it Initial states.} The initial state of \(L-1\) unmeasured qubits are considered to be the thermal state of their local Hamiltonian, 
 \begin{eqnarray}
 H_{A_{i}} = \frac{\omega_{A_{i}}}{2} \sigma_i^z,
 \label{eq:localham}
 \end{eqnarray}
  with $\sigma^{\mu}_{i}$ ($\mu = x, y, z$) is the Pauli matrices acting on the subsystem, \(i\) while the measured one is initially prepared as the ground state of the same Hamiltonian. Here $\omega_{A_{i}}$ is the strength of the external magnetic field on the qubit $A_{i}$. Therefore, the initial state reads as
  \begin{equation}
      \rho_{in} = \bigotimes_{i\ne j} \frac{1}{\mathcal{Z}_i}\exp(-\beta H_{A_{i}})\otimes |\mathcal{G}_{A^{M}_j} \rangle\langle \mathcal{G}_{A^{M}_j}|,
  \end{equation}
  where $\beta=\frac{1}{k_B T}$ is the inverse temperature corresponding to the absolute temperature, $T$, with Boltzmann constant, \(k_B\) and $\mathcal{Z}_i=\Tr[\exp(-\beta H_{A_{i}})]$. We assume $\beta$ to be same for all the unmeasured spins. 

{\it Evolution operator.} The integral part of a refrigerator is the unitary evolution of the whole system for establishing the quantum  correlation between the qubits which plays an important role to achieve cooling.   The unitary dynamics is dictated by the local Hamiltonian  $H_{A_{i}}$ ($i = 1, 2 \ldots L$) along with the interacting variable range \(XY\) Hamiltonian, 
\begin{eqnarray}
H_{int} = \sum_{\substack{i < j}}^{L} \frac{J_{ij}}{4}\Big[(1+\gamma)&\sigma_i^x&\sigma_j^x + (1-\gamma)\sigma_i^y\sigma_j^y \Big],
\label{eq:ham1} 
\end{eqnarray}
where $\gamma$ is the anisotropy parameter in the $xy$ plane, $J_{ij}$ is the interaction strength between the qubit $i$ and $j$, which varies according to the variable range interaction, $J_{ij} \sim \frac{J}{|i-j|^{\alpha}}$, with $\alpha$ being the fall-off rate of the interaction strengths between the different subsystems and $J$ is a coupling constant which dictates ferromagnetic $(J < 0)$ or anti-ferromagnetic $(J > 0)$ type interaction between the qubits. Note that large value of $\alpha$ i.e., $\alpha > 2$ actually mimics the nearest neighbor Hamiltonian while $\alpha \leq 1$ mimics the long-range behavior. Therefore, the Hamiltonian, which dictates the evolution can be written as \(H_{ev} = \sum_i H_{A_i} + H_{int}\). 

{\it Measurements.} Let us consider a linear array of spins in which projective measurements  $\{\Pi_{k}^{M} = |\mathcal{G}_{A_{j}^{M}}\rangle, |\mathcal{E}_{A_{j}^{M}}\rangle\}$ is performed on the qubit $A_{j}^{M}$. In this scenario, $\{\Pi_{k}^{M}\}$ reduces to a measurement in the computational basis, i.e., $\{|0\rangle, |1\rangle\}$ at each step. We continue the process only when the outcome $\ket{1}$ which is the ground state of the local Hamiltonian in Eq. (\ref{eq:localham}) clicks.

\section{Three qubit measurement-based Refrigerator }
\label{sec:ana}

Let us first illustrate  the performance of the measurement-based refrigerator, consisting of three spin-$\frac{1}{2}$ particles. The refrigerators built with higher number of qubits will be dealt in the succeeding section.

{\it First qubit as the accessible one.} Let us consider the scenario where the qubits are placed in a 1D array (as in Fig. \ref{fig:fig1}) and the measured one is the first qubit, $A^M_1$ while the rest of the qubits are denoted by \(A_2\) and \(A_3\). The initial state in this case  is taken as 
\begin{eqnarray}
\rho(0) = |\mathcal{G}_{A^{m}_1} \rangle \langle \mathcal{G}_{A^{m}_1}| \bigotimes_{i=2}^{3}  \frac{1}{\mathcal{Z}_i}[\exp(-\beta H_{A_{i}})],
    \label{eq:initial_state3}
\end{eqnarray}
where the unmeasured qubits are  the thermal states of the local Hamiltonian in Eq. (\ref{eq:localham}) and the measured qubit is the ground state of the same Hamiltonian.  We will argue later that  the spatial position of the measured qubit also effects the cooling mechanism. 
Notice that when the temperatures of the unmeasured initial states are below some critical temperature, the local states are  close to the ground state of the local Hamiltonian, \(H_{A_{i}}\), thereby leading to a trivial situation. On the other hand, the interesting scenario emerges when the temperature of the initial state is high enough.

Let us start with an extreme case, i.e., when the unmeasured qubits are initially prepared as maximally mixed, i.e., the thermal states with \(\beta =0\). The system evolves according to the nearest neighbor \(XX\) Hamiltonian in presence of magnetic field, given in Eqs. (\ref{eq:ham1}) and (\ref{eq:localham}) with \(\gamma =0\), \(\alpha > 2\), \(\omega_{A_i} =1\) while the measurement on the qubit \(A^M_1\) is performed in the basis of the local Hamiltonian, i.e., \(\{|\mathcal{G}_{A^{M}_1}\rangle, |\mathcal{E}_{A^{M}_1}\rangle \}\). 
All the qubits evolve for a short period of time, $t$ which is followed by a measurement. After \(N\) steps of the evolution and measurement, the resulting tripartite state can be found to be
\begin{eqnarray}
\rho(t) & = & x |100\rangle \langle 100| + y P[a |110\rangle + b |101\rangle] \nonumber \\
&+& (1-y-x) |111\rangle \langle 111|,
\label{eq.10}
\end{eqnarray}
where  \(|1\rangle\) and \(\ket{0}\) represent the ground and the excited  states of  \(H_{A_i}\) respectively, \(x, y, a\), and \( b\) depend on the  system parameters as well as repetition of the process, $N$ and \(P\) is the projector. When the outcome of the measurement is \(|\mathcal{G}_{A^{M}_1}\rangle\), the output state at each subsystem, $\rho_{A_i}^f$ turns out to be  diagonal in the computational basis, given by 
\begin{eqnarray}
\rho^f_{A_{i}} = \frac{1}{\mathbb{N}} (\eta_{A_{i}} | \mathcal{G}_{A_{i}} \rangle \langle  \mathcal{G}_{A_{i}}| + \eta'_{A_{i}}| \mathcal{E}_{A_{i}} \rangle \langle  \mathcal{E}_{A_{i}}|).
\end{eqnarray}
Here \(|\mathcal{E}_{A_{i}} \rangle\) denotes the excited state of the local Hamiltonian, the coefficients,  \(\eta_{A_{i}}\) (\(\eta'_{A_{i}}\)) are  functions of \(\sum_{r=1, 3, \ldots}^{2N-1} A_r \cos(\frac{J t r}{\sqrt{2}})\) with only odd \(r\), and   \(\sum_{r=1}^N B_r \cos(\sqrt{2}J t r)\) upto some  constant factor and \(\mathbb{N}\) is the normalization constant. 
To quantify the performance of the refrigerator, the corresponding fidelity of the qubit, \(A_i\) (\(i = 2, 3\)),  can be represented as
\begin{eqnarray}
\mathcal{F}_{A_i}(N) = \langle \mathcal{G}_{A_{i}} | \rho_{A_i}^f |\mathcal{G}_{A_{i}}\rangle,
\end{eqnarray}
which  takes the form after \(N\) rounds as
\begin{eqnarray}
\mathcal{F}_{A_i} (N) = \frac{1}{\mathbb{N}}\Big[C^N_{A_{i}} &+& \sum_{r=1,3 ... }^{2N-1} C'^r_{A_{i}} \cos (\frac{J r t}{\sqrt{2}})  \\ \nonumber
&+& \sum_{r=1}^{N}C''^r_{A_{i}} \cos (\sqrt{2} J t r)\Big], \,\, i=2, 3.
\label{eq:10}
\end{eqnarray}
The coefficients, \(C^N_{A_{i}}\), \(C'^r_{A_{i}}\) and \(C''^r_{A_{i}}\) depend on the number of times evolution and measurement operators, \(N\)  acted on the system and other system parameters.

 \begin{figure}
     \centering
     \includegraphics[height=5 cm,width=7 cm]{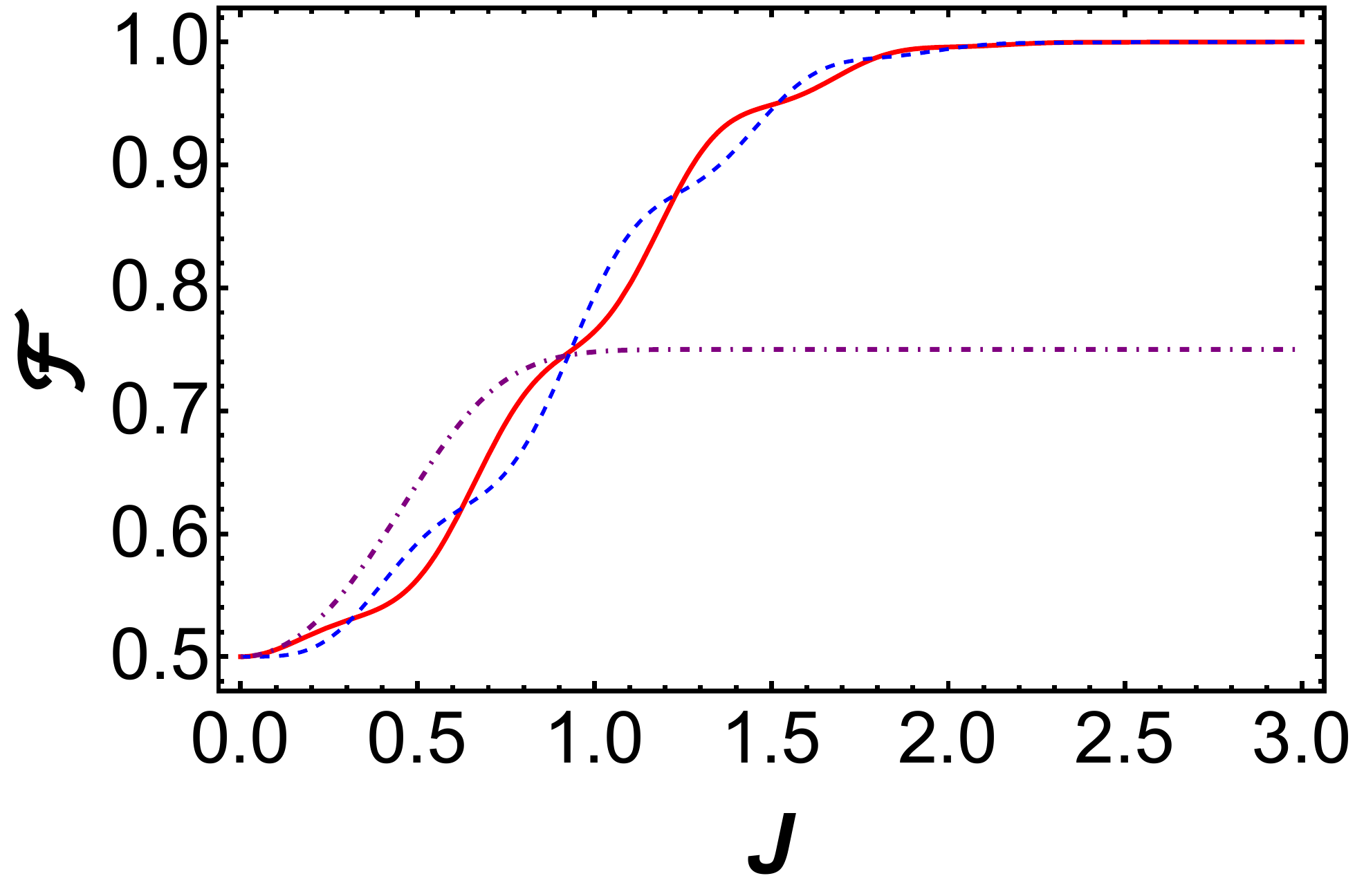}
     \caption{(Color online) \textbf{ Variation of fidelity as a function of interaction strength $J$}. The refrigerator consists of three qubits which are interacting with each other via nearest neighbor $XX$ Hamiltonian in presence of magnetic field in Eq. (\ref{eq:ham1}) with \(\alpha > 2.5\). The initial state is a fully separable state, given in Eq. (\ref{eq:initial_state3}) with \(\beta =0\). Dashed  and solid lines represent the fidelities of the second and the third qubits (ordinate) with respect to \(J\) (abscissa) when the measurement is performed in the first qubit. The dashed-dotted line correspond to  \(\mathcal{F}\) for the first and the third qubits due to the symmetry with the measurement being done on the second qubit.      Here $t=1$, $\omega_i^{A_i}=1$, $\forall i \in \{1,2,3\}$. Both  the axes are dimensionless.  }
     \label{fig:fid3qubit}
 \end{figure}
After a few such steps, the fidelity goes close to  unity  provided \(Jt\) is chosen appropriately. To visualize the dependence on parameters,  we plot $\mathcal{F}_{A_2}$ and $\mathcal{F}_{A_3}$ by varying $J$ for $N=10$ and $t=1$ in Fig. \ref{fig:fid3qubit}. 
Notice that the two-party reduced density matrix  between the inaccessible qubits which is obtained after measuring on the first qubit,  is  again a function of $x$, $y$, $a$ and $b$ and the second term is responsible for generating entanglement in the system. We find that by increasing the number of measurements, $N$,  the cross-term vanishes, which leads to a fully separable state
of their corresponding ground states (i.e., $\mathcal{F}_{A_{i}} = 1$), describing the phenomena of cooling. Due to  symmetry, the similar cooling phenomena occurs when the measurement is performed on the third qubit.

\emph{Second qubit as the measured qubit.} Suppose the measured qubit is the second one, i.e., the system is initially prepared as
\begin{eqnarray}
\rho(0) = \frac{1}{\mathcal{Z}_1}\exp(-\beta H_{A_1}) \otimes |\mathcal{G}_{A^{M}_2} \rangle \langle \mathcal{G}_{A^{M}_2}| \otimes   \frac{1}{\mathcal{Z}_3}\exp(-\beta H_{A_3}), \nonumber
\end{eqnarray}
with \(\beta =0\).
After evolving $\rho(0)$ via \(XX\) Hamiltonian with a magnetic field, \(H_{ev}\), and measuring the middle qubit in the $\{\Pi_k^M\}$ basis, we find that  both the first and the third qubits reach to the same diagonal  states when $\ket{1}$ clicks which is in contrast to  the previous situation.  Since both the first and the third qubits are the nearest neighbor with the measured qubit, the system possess a symmetry which finally leads to the same state at the first and third subsystems after \(N\) steps. In this case, after \(N\) repetitions, the output three qubit state becomes
\begin{eqnarray}
\rho(t) & = & x' |010\rangle \langle 010| + y' P[a' |110\rangle + b' |011\rangle] \nonumber \\
&+& (1-y'-x') |111\rangle \langle 111|,
\end{eqnarray}
and the corresponding 
 fidelity in any unmeasured subsystem is given by
\begin{eqnarray}
\mathcal{F}(N) = \frac{4^{N-1}}{C^N + \sum_{r=1}^N C'^r \cos(\sqrt{2}J t r)},
\label{eq:12}
\end{eqnarray}
where \(C^N\) and \(C'^r\) takes the values depending on the number of rounds evolution and  measurements being carried out. Comparing fidelities in Eqs. (\ref{eq:10}) and (\ref{eq:12}), we observe that fidelities depend on the spatial position of the unmeasured qubit with respect to the measured one. Unlike the previous scenario,  the entanglement generating term in this case does not vanish  with the increase of $N$. As a result, by measuring on the second qubit, we can not achieve refrigeration since the joint state of the unmeasured qubits  always remains entangled. Later we will discuss the dependence of spatial position of the measured qubit towards the refrigeration in  details.  Note, however, that the probability is higher for the case having non-unit fidelity than the one with unit fidelity.

 \begin{figure*}
     \centering
     \includegraphics[height=6 cm,width=17 cm]{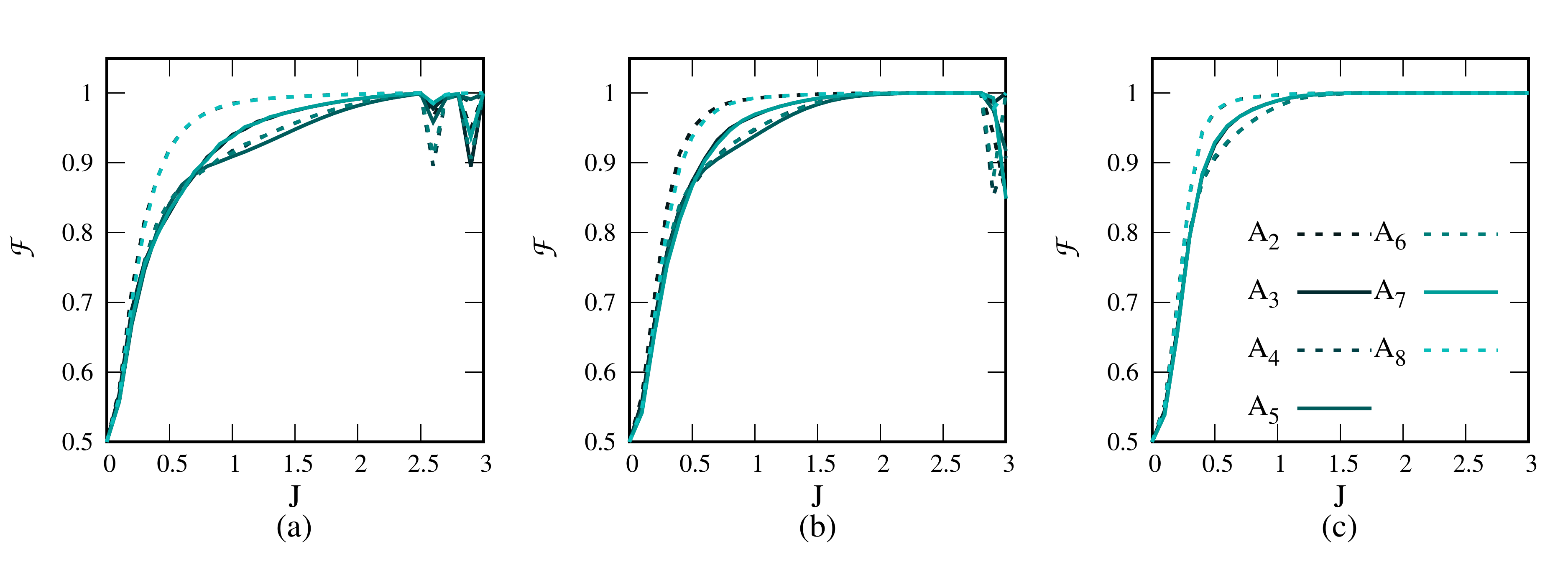}
     \caption{(Color Online.) \textbf{Fidelity (ordinate) against interaction strength between qubits, $J$ (abscissa) for different values of $\alpha$.} Here (a), (b) and (c) correspond to $\alpha = 1.0, 1.5$ and $2.5$ respectively. 
     The initial state is given in Eq. (\ref{eq:initial_state_infinite}) with \(\beta =0\) and the evolution Hamiltonian contains both the interacting Hamiltonian in Eq. (\ref{eq:ham1}) with \(\gamma =0\) and local Hamiltonian, \(H_{A_i}\) in Eq.  (\ref{eq:localham}). \(A_i\)s represent the unmeasured qubits while the measurement is performed on the first (last) qubit of the chain  and the outcome of the measurement is the ground state of the local Hamiltonian. Dashed (solid) lines represent even (odd) positions of the qubits. In both the cases, dark to lighter shades occur with the increasing value of the position of the qubits. System size is taken as $L=8$. Here $N=500$, $\omega_i^{A_i}=1$, $\forall i \in \{1,\ldots 8\}$. All the axes are dimensionless. }
     \label{fig:fig2}
 \end{figure*}

    \begin{figure*}
     \centering
     \includegraphics[height=6 cm,width=17 cm]{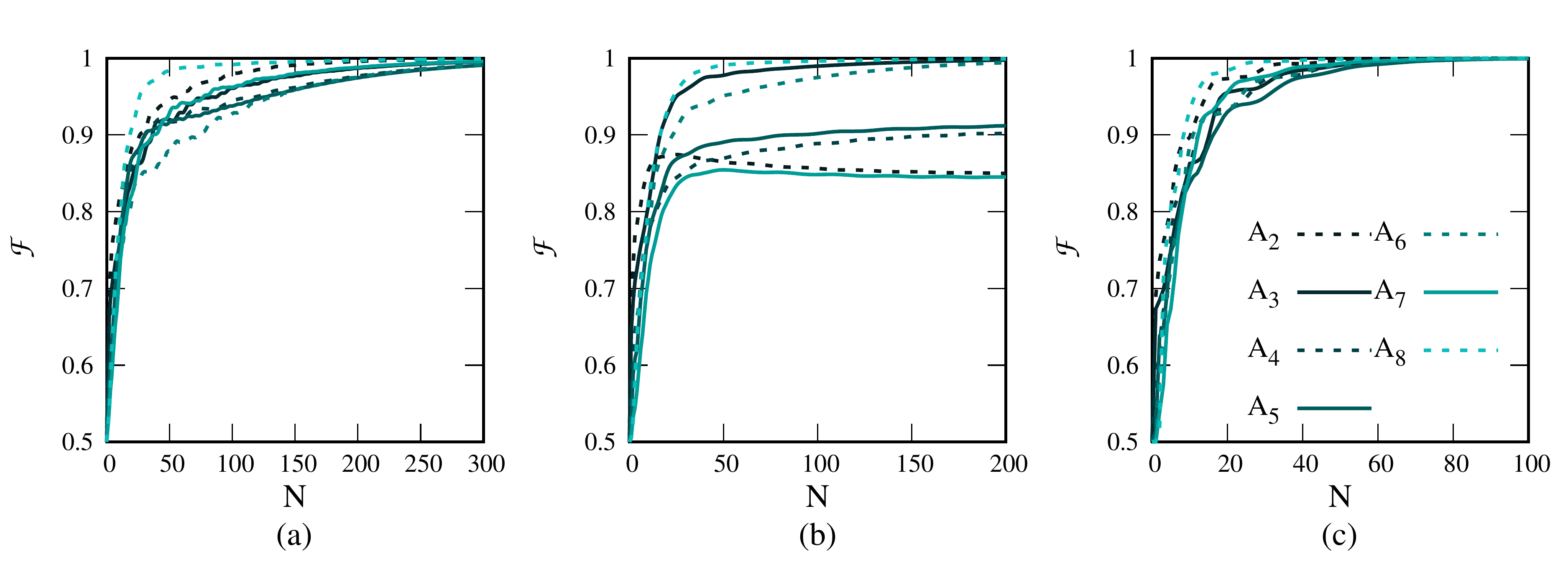}
     \caption{(Color Online.) \textbf{Fidelity (vertical axis) with respect to number of rounds, $N$ (horizontal axis).} Here interaction strength, $J =3$. 
     All other specifications are same as in Fig. \ref{fig:fig2}. 
     All the axes are dimensionless.
     }
     \label{fig:fig3}
 \end{figure*}

\section{Refrigeration with arbitrary number of sites}
\label{sec:refrimorequbits}

We will now investigate the performance of a refrigerator which is comprised of  arbitrary number of qubits.  
We consider a set-up when all the unmeasured qubits are initially prepared at infinite temperature, i.e., the states are maximally mixed and the measured qubit is in the ground state of its corresponding Hamiltonian. Moreover, we assure here that the spins are situated in a chain with open boundary condition and the first or the last qubit is chosen for measurement. Without loss of generality, the initial fully separable state of the total system 
reads as
\begin{equation}
    \rho(0) = |\mathcal{G}_{A^{M}_1} \rangle \langle \mathcal{G}_{A^{M}_1}| \bigotimes_{i=2}^{L} \big[\frac{1}{2} (|0\rangle \langle 0| +|1\rangle \langle 1|)\big].
    \label{eq:initial_state_infinite}
\end{equation}
When $t>0$, the interaction between the qubits is turned on which allows the evolution of the total system according to $\mathcal{U}(t) = \exp(- i H_{ev} t)$.
Our aim is to observe the final states of the unmeasured qubits after $N$ rounds of projective measurements, where one of the elements can be represented as  $\Pi_1 = |\mathcal{G}_{A^{M}_1}\rangle\langle \mathcal{G}_{A^{M}_1}| \bigotimes \mathbb{I}_{L-1} = \ket{1}\bra{1} \bigotimes \mathbb{I}_{L-1}$, with \(\mathbb{I}_{L-1}\) being the identity operator of the unmeasured Hilbert space. The phenomena of refrigeration demands the situation when the states of all the unmeasured qubits are in the ground state of the respective Hamiltonian, i.e., when the projection operator, $\ket{1}$ clicks.   

As we have shown in case of three qubits in  Fig. \ref{fig:fid3qubit},    system parameters involved in the process play a crucial and delicate role in this cooling procedure. Let us first explain the choice of the system parameters which are relevant during dynamics for successful implementation of a quantum refrigerator with arbitrary number of sites. Numerical simulation is performed of a 1D array containing \(L=8\) sites. 

\begin{figure*}
     \centering
     \includegraphics[width=\textwidth]{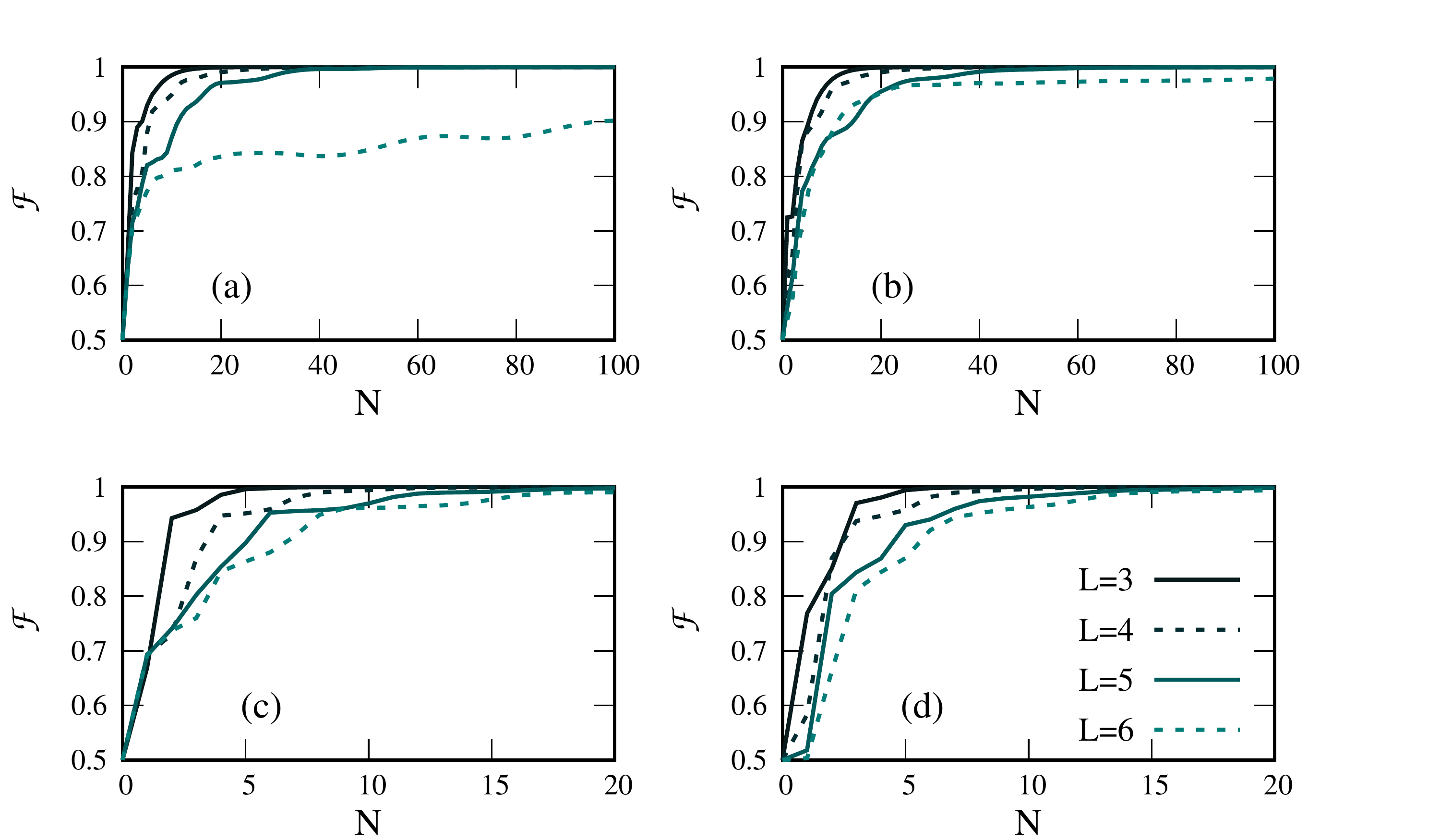}
     \caption{(Color Online.) \textbf{Fidelity (ordinate) as a function of number of measurements, $N$ (abscissa) for different system sizes, $L$.} We study the dependency of fidelity of the unmeasured qubit adjacent to the measured qubit and the qubit situated farthest from the measured one. 
     Solid and dashed lines represent odd and even lattice sites respectively. 
     Other parameters of the systems are $J=3$ and for  (a) and (b), $\alpha=1$ and for (c) and (d), $\alpha=2.5$.  (a) and (c) correspond to the fidelity for qubit adjacent to a measured qubit, i.e., the second qubit since the measurement is carried out on the first qubit while (b) and (d) represent the behavior of \(\mathcal{F}\) for the farthest qubit, i.e. the qubit in the boundary, \(A_i\) (\(i=3, 
     \ldots 6\)). All other specifications are same as in Fig. \ref{fig:fig2}. All the axes are dimensionless.
     }
     \label{fig:fig4}
 \end{figure*}


\emph{Effect of interaction strength.} Let us take the initial state as in Eq. (\ref{eq:initial_state_infinite}) and $H_{ev}=H_{A_i}(\omega_{A_i}=1)+H_{int}(\gamma=0)$. We also fix the number of repeated evolution and measurement to be $N=500$ and the time interval in which the evolution takes place is chosen to be unity. In this set-up, we observe that the fidelity between the  reduced subsystem, \(A_i\) and the respective ground state of that particular qubit increases with the increase of interaction strength, $J$ as depicted in Fig. \ref{fig:fig2} for different values of $\alpha$, with $L=8$. It implies that reasonably high interaction strength is required for a successful implementation of a refrigerator. In addition,  we observe that  ferromagnetic or antiferromagnetic interactions do not play a role in the performance which we can also understand from the expressions of fidelities obtained in Eqs. (\ref{eq:10}) and (\ref{eq:12}).

Moreover, for nearest-neighbor Hamiltonian (i.e., when \(\alpha =2.5\)), all the unmeasured qubits reach to their corresponding ground state for small value of $J$ compared to the evolution governed by the long-range quantum spin models (i.e., \(\alpha <2\)). Notice that the moderate values of $N \geq 100$, \(L\) and $t \geq 1$ do not alter the results qualitatively. However, the  presence of anisotropy in the dynamics typically has detrimental effect on fidelity as well as probability, thereby establishing that the  $XX$ interacting Hamiltonian leads to the best performance in cooling.

Interestingly, the fidelity of the subsystem follows a particular pattern with $J$ in a spin chain of length $L$.  If the measurement is carried out on the first qubit, for a chain with even \(L\), the qubit $(k+1)$ and  $(L-k+1)$ ($k = 1, 2, 3, \ldots (\frac{L}{2} - 1)$)  and for odd \(L\), $\lfloor \frac{L}{2} \rfloor$  reach the maximum fidelity simultaneously with $J$ for a fixed $N$. This is due to the inherent parity-symmetry present in the system even when the open boundary condition is taken. Strikingly, the farthest and the nearest qubit from the measured one reach the unit fidelity with less number of repeated measurements, $N$, compared to all other qubits located in the middle of the spin chain  which require high values of $N$ and interaction strength to achieve the same.

\begin{figure}
     \centering
     \includegraphics[scale=0.6]{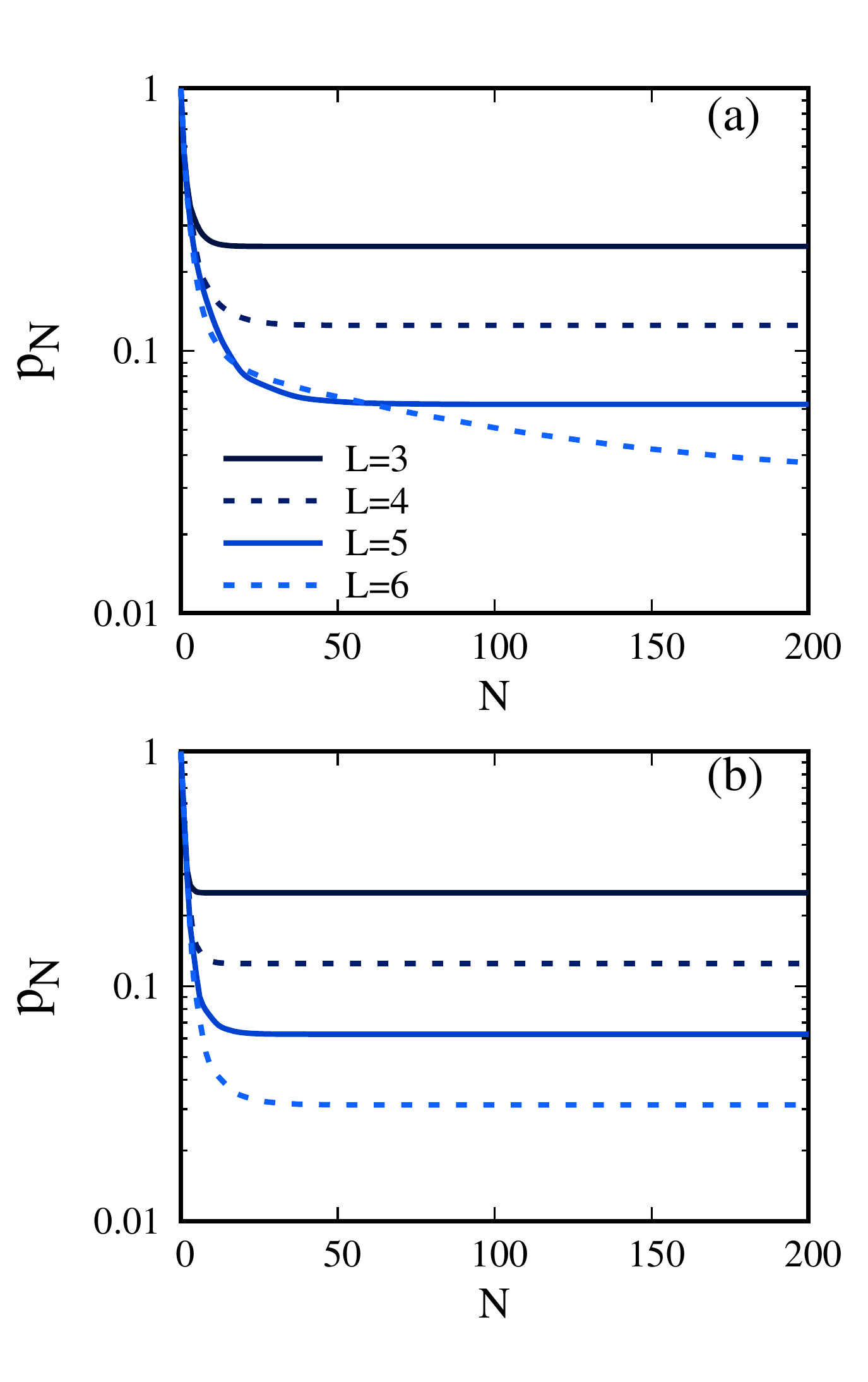}
     \caption{(Color Online.) \textbf{Probability \(p\) in log scale (\(y\)-axis) vs. $N$ (\(x\)-axis) for different system sizes.} 
     (a) \(\alpha =1\) and (b) \(\alpha =2.5\). Here \(J=3\) in \(H_{ev}\).  All other specifications are same as in Fig. \ref{fig:fig2}. Irrespective of system sizes,  we find that the success probability to achieve cooling saturates to a nonvanishing value, thereby ensuring the usefulness of this procedure. All the axes are dimensionless.  }
     \label{fig:fig5}
 \end{figure}

\emph{Dependence on number of measurements.}   To investigate the number of steps required to cool down the subsystem of a spin chain of length $L$, we fix the other parameters. It is important to stress here that although some specific parameters are chosen for presenting the results, it is possible to  identify the parameter regime where cooling phenomenon can be found (as shown in Fig.  \ref{fig:fig2}). We have already realised that  successful realization of a refrigerator depends on the strong interaction strength in the dynamics and hence for demonstration, we choose $J=3$, $\omega_{A_i}=1$, $t=1$ and $L=8$. The results presented here remain unaltered with other system parameters like $J>3$, $t>1$, starting with the same initial state in Eq. (\ref{eq:initial_state_infinite}). Moreover,  we tune the fall-off rate of interactions between the qubits so that we can generate the effects of nearest-neighbor as well long-range interactions on fidelities. Let us summarize the observations.

\begin{enumerate}
    \item  After evolution and measurements as per prescription, we find that  all the unmeasured qubits can be projected to their ground state for different values of variable range interaction strength (see  Fig. \ref{fig:fig3}). Notice that in Fig. \ref{fig:fig3}, the choice of $J$ turns to be suitable for $\alpha=1$ and $\alpha=2.5$ although  it is not appropriate for $\alpha=1.5$ and hence there are qubits which do not approach to the unit fidelity. However, proper tuning of \(J\)  can also induce fidelities of all the qubits to unity even for \(\alpha =1.5\) as seen in Fig. \ref{fig:fig2}.  

    \item An interesting question can be addressed at this point --  does the distance of a qubit from the measured one matter to achieve the maximum fidelity of a qubit? More specifically, we figure out the position of the qubit that reaches the maximum fidelity with minimum $N$. We observe that irrespective of $\alpha$, most distant qubit  from the measured qubit goes to the ground state with minimum number of measurements than any other qubits present in the system. However,  $N$ required for other unmeasured qubits do not follow any hierarchies based on the distance from the measured qubit.  
    
    
    
\end{enumerate}

\subsubsection*{Scalability -- Role of  system size}  

Let us address here the scalability of the machine.  In other words, when one increases the system size and measures only on a single qubit repeatedly, we investigate the scaling of fidelity with $L$. Since the number of subsystems to be cooled increases with $L$, we notice that number of measurements require to cool  also increases with $L$ as shown  in Fig. \ref{fig:fig4} for two different values of \(\alpha\).  Moreover, we observe that for a fixed $L$ with high values of $\alpha$, i.e., when the interaction strength is confined between the nearest pair of qubits, it is possible to decrease the temperature of all the qubits much more efficiently than the model with long-range interactions, comparing (a), and (b) with (c) and (d) of Fig. \ref{fig:fig4}. 

When the number of measurements are low enough, there is no prominent pattern of fidelity with system size. 
However, for moderate values of $N$, clear pattern with increasing system size emerges as depicted in Fig. \ref{fig:fig4}.


In addition, we  also examine the success probability of obtaining the measured qubit in the corresponding ground state after  completion of each  round. From Fig. \ref{fig:fig5}, it is clear that increasing system size decreases the success probability monotonically for a moderate $N$. It is also independent of the range of  interactions.  Nonetheless, the success probability saturates with increasing number of measurements, thereby establishing the usefulness of this protocol. With $L$, the success probability follows a scaling law which we will discuss below.

 \begin{figure}
     \centering
     \includegraphics[height=6 cm,width=9 cm]{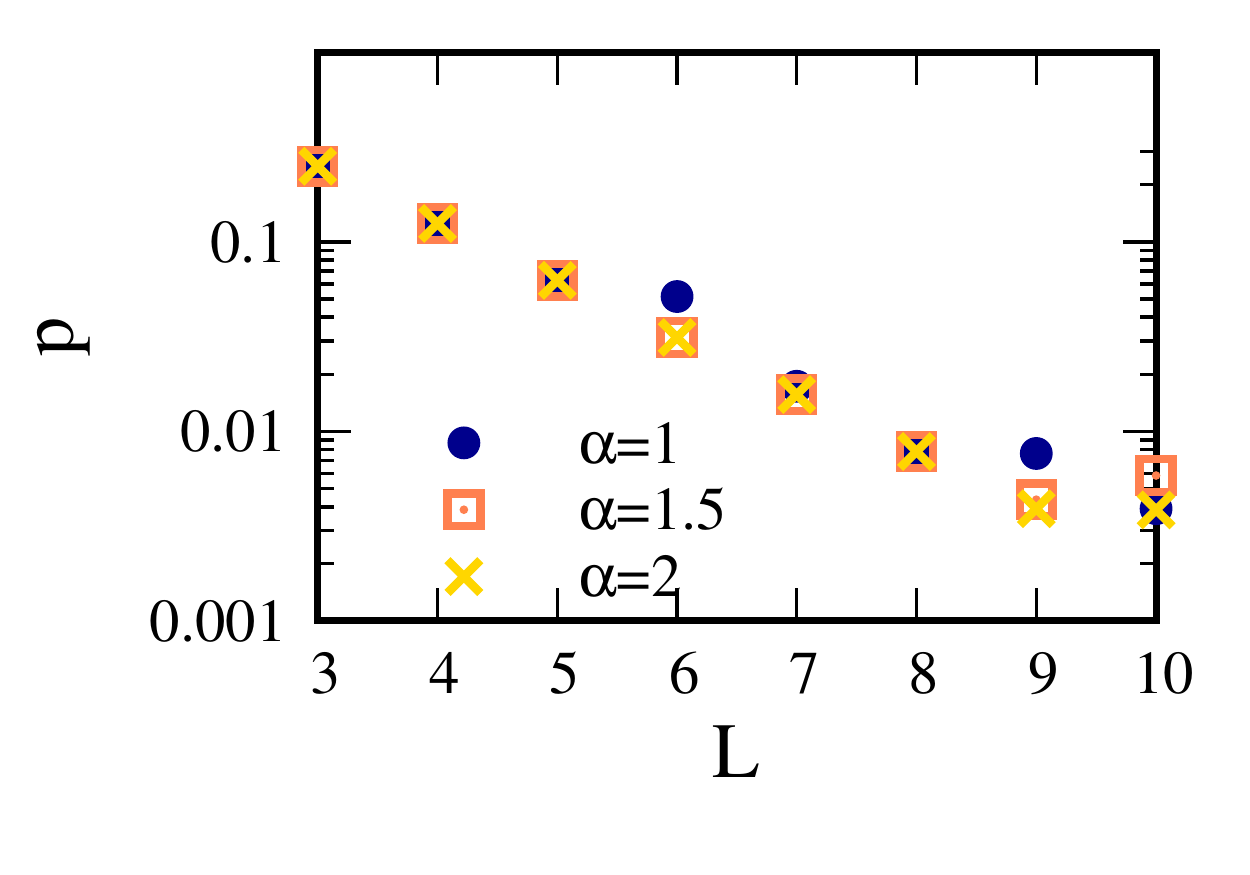}
     \caption{(Color Online.) {\bf Probability in log scale (ordinate) against system size (abscissa). } Circles, squares and crosses represent different variable range of interactions. It depicts that \(p\) decreases with the increase of \(L\). However, it is independent of the fall-off rate of interactions, \(\alpha\). All other specifications like initial state, evolution operators are same as in Figs.
     \ref{fig:fig2} and \ref{fig:fig3}. Both the axes are dimensionless. }
     \label{fig:fig6}
 \end{figure}

\subsection{Scaling}
\label{sec:scaling}

 \begin{figure*}
     \centering
     \includegraphics[width=\textwidth]{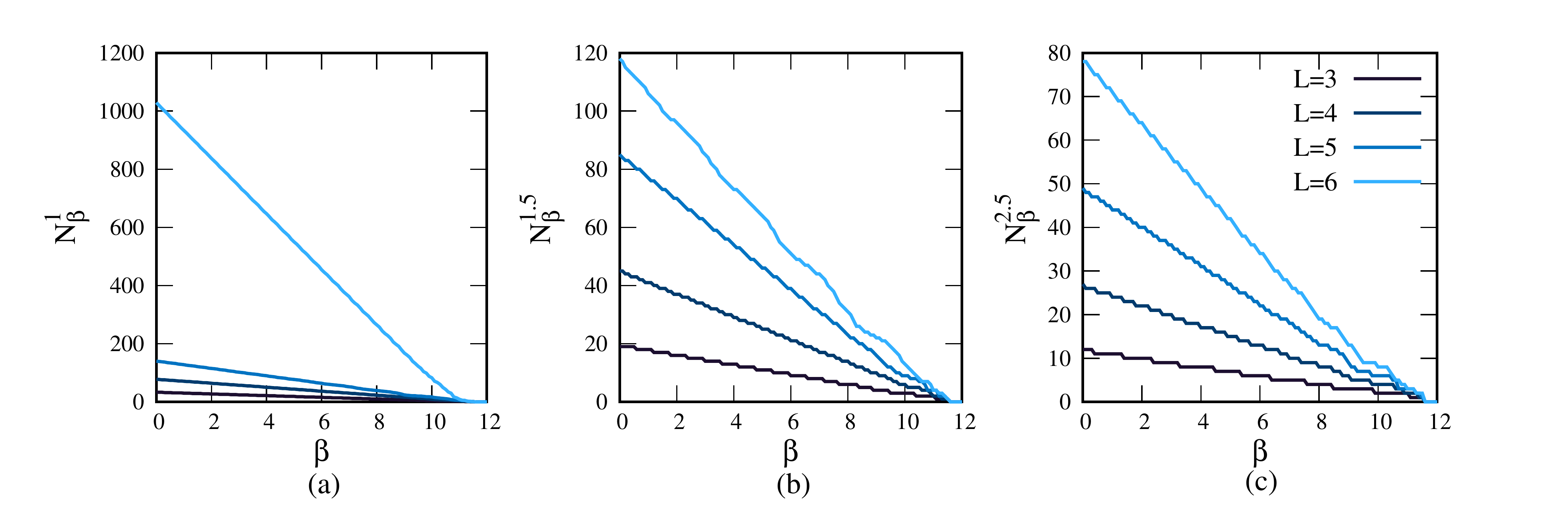}
     \caption{(Color Online.) \textbf{Maximum number of measurements $N_{\beta}^{\alpha_i}$ required to obtain the unit fidelity in all the subsystems (vertical axis) with the initial inverse temperature, $\beta$ (horizontal axis).} Here (a) \(\alpha_1 =1\), (b) \(\alpha_2 = 1.5\) and (c) \(\alpha_3 =2.5\).  Different lines correspond to different system sizes, $L$.   The initial state is the thermal state with nonvanishing \(\beta\). Dark to lighter shades occur with the increase of L. All other parameters are same as in Fig. \ref{fig:fig2}. All the axes are dimensionless.} 
     
     \label{fig:fig7}
\end{figure*}

Intuitively, in a measurement-based protocol, the success probability of finding the measured qubit in its corresponding ground state decreases with the increase of number of measurements. However, we find out that $p_{N}$ first decreases and finally saturates after a certain value of $N$ as seen in Fig. \ref{fig:fig5}. When the fidelities of all the individual unmeasured qubit approach to the maximum fidelity i.e.,  \(\mathcal{F}=1\) after some steps, \(N\), we observe that the corresponding probability follows a scaling law with the system size, $L$, given as
\begin{equation}
    p=\big(\frac{1}{2}\big)^{L-1},
    \label{eq:probscale}
\end{equation}
which is independent of all the parameter values and the number of measurements. $L$ is the total number of spin-$\frac{1}{2}$ particles in the spin chain. On the other hand, no such kind of general relationship between the fidelity of the individual qubits and the system size is observed.
Numerically, the scaling law of $p$ for different values of 
 \(\alpha\) is depicted in Fig. \ref{fig:fig6} which is in good agreement with the expression in Eq. (\ref{eq:probscale}). 

\subsection{Dependence of initial temperature}
\label{sec:initial_temp_dependence}

Upto now, we consider the scenarios where  the unmeasured qubits are prepared in  maximally mixed states i.e., the thermal states with $\beta = 0$. Let us deal with the situation where the unmeasured initial states are the thermal state  with a moderate temperature. Such an analysis can reveal the dependence of initial temperatures of unmeasured subsystems on the phenomena of refrigeration.

The minimum number of  measurements required for attaining the ground state of all the individual qubit  decreases with the increment of inverse temperature of the initial states, i.e., with the decrease of temperature provided the involved system parameters are chosen appropriately. This behavior can be intuitively explained since the increasing value of $\beta$ of an individual qubit makes the subsystem close to their respective ground state and hence less number of measurements are sufficient to cool down the system. However, the pattern of reduction of $N$ with \(\beta\) is not apriori fixed.  As observed in Fig. \ref{fig:fig7},  the maximum number of measurements among all the unmeasured qubits is a linear function of $\beta$. 

Let us define $N^{\alpha_i}_\beta$ as the number of steps required to reach the unit fidelity for all the unmeasured qubits. We observe that for moderate values of $\alpha>1$, $N^{\alpha_i}_\beta$ scales linearly with $\beta$ and it increases monotonically with the system size. On the other hand, for $\alpha \leq 1$, and for a fixed $L$, $N^{\alpha_i}_\beta$ is much higher than the one with $\alpha>1$ although the linear behavior of $N^{\alpha_i}_\beta$ with inverse temperature still holds.
It exhibits that although the interaction between the subsystems which can create correlation between the measured qubit are required to achieve cooling, the role of interaction on cooling is not ubiquitous. Precisely, for fixed value of $\beta$, we find that 
\begin{equation*}
N^{\alpha_{1}}_{\beta} < N^{\alpha_{2}}_{\beta}< \ldots 
\end{equation*}
where  $\alpha_{1} > \alpha_{2} > \ldots $. It indicates that irrespective of temperature of the initial state, range of interactions and interaction strength both have important role to play in refrigeration.


\section{Spatial position  of measured qubits vs. entanglement distribution of unmeasured qubits }
\label{sec:spatial}

 Due to   repeated measurements on the accessible qubit located in the boundary of the spin chain, and by evolving the entire system according to the interacting Hamiltonian, we have already shown that the local energy for the rest of the system can be minimized provided suitable parameters are chosen. Removing the constraint on the position of the measured qubit, we explore here the situation when the measurement can be carried out on any arbitrary qubit placed at an arbitrary position in an 1D array. Specifically, once the position of the measured qubit is fixed, all the rounds of the protocol that particular qubit acts as the \emph{measured qubit}. Interestingly, we find out a notable difference between this and the aforementioned scenario.
 
A refrigerator is built of a $L$ spin-$\frac{1}{2}$ particles arranged in a liner chain with the initial state being
 \begin{equation}
     \rho(0) = | \mathcal{G}_{A_{r}^{M}} \rangle \langle \mathcal{G}_{A_{r}^{M}} | \bigotimes \rho_{in}^{L-1},
 \end{equation}
 where $\rho_{in}^{L-1}$ is the thermal state of individual local Hamiltonian, $H_{A_{i}}$ having $\omega_{A_{i}} = 1$ and $\beta = 0$. Here the site on which the measurement is performed is taken to be $A_{r}^{M}$. After the evolution according to  $H_{ev}$ and the outcome being $|\mathcal{G}_{A_{r}^{M}} \rangle$, the output state after $N$ rounds, becomes 
 \begin{eqnarray}
      \rho(t) &=& \nonumber x_{1}|0 \ldots 1 0 \ldots \rangle \langle 0 \ldots 1 0 \ldots 0 | \\&& \nonumber + x_{2} P_{2} [\sum a_{r}^{1} \mathcal{P}[0 \ldots 1 1 \ldots 0]] \\&& \nonumber + x_{3} P_{3} [\sum a_{r}^{2} \mathcal{P}[0 \ldots 1 1 1 \ldots 0]] + \ldots  \\&& +  x_{L} |1 1 1 \ldots 1 \rangle \langle 1 1 \ldots 1|,
 \end{eqnarray}
 where $\mathcal{P}[0 \ldots 1 1 0 \ldots 0]$ denotes the permutation operator which permutes $(L-1)$ number of $|1\rangle$ in all the $L$ places, keeping $|1\rangle$ fixed to the position $A^{M}_{r}$, on which the measurement is performed. The position of \(\ket{1}\) in the first term is at the \(A_r^M\) position, in other terms,  \(P_i\)s denote the projectors and subscripts denote the number of $|1\rangle$ present in the projector. The coefficients, \(x_i, a_r^k\)s are functions of system parameters. 
 \begin{figure}
     \centering
     \includegraphics[height=4.5 cm,width=9 cm]{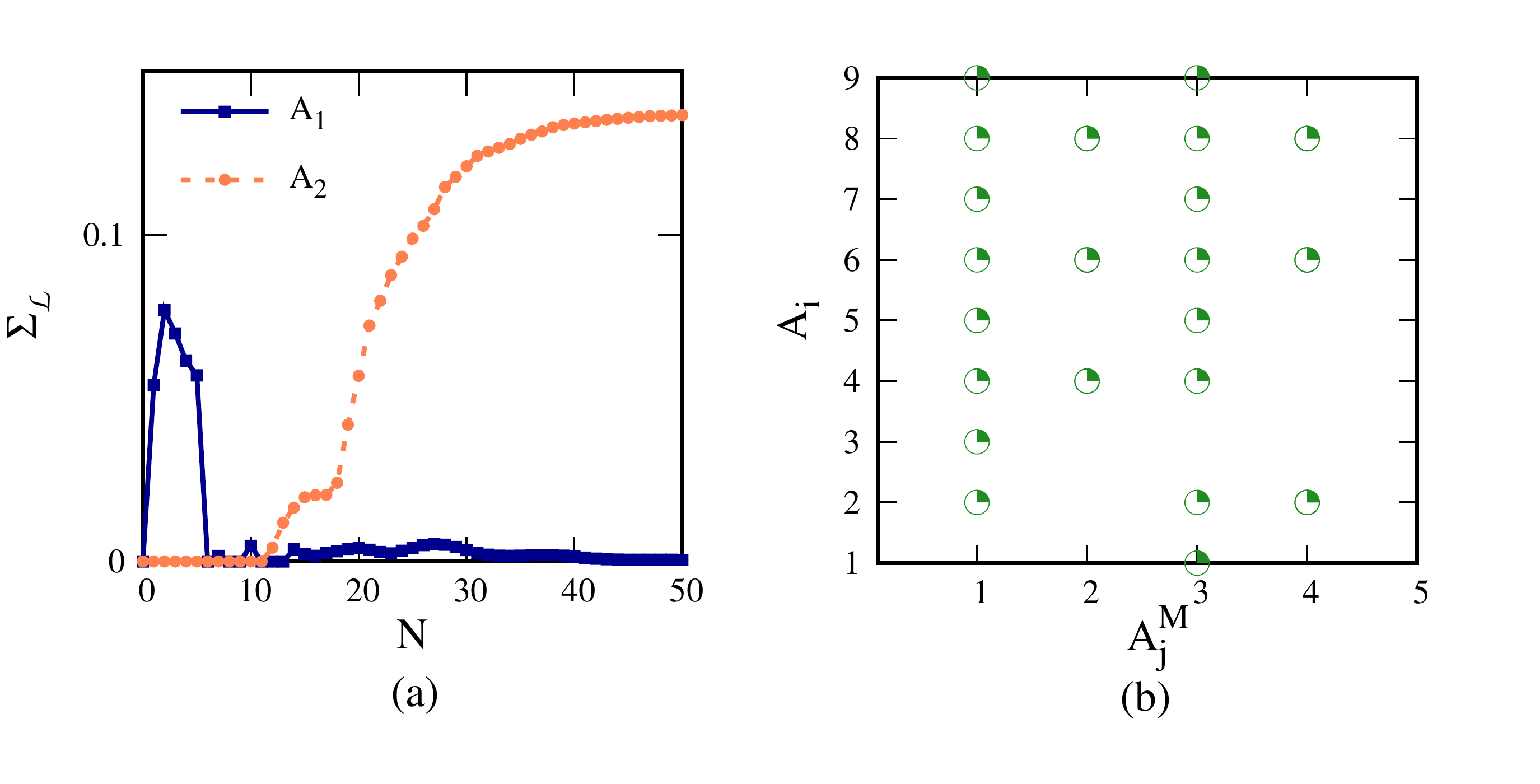}
     \caption{{\bf Distribution of entanglement of unmeasured qubits and role of position in a measured qubit.} (a) Distribution of entanglement, \(\Sigma_{\mathcal{L}}\) defined in Eq. (\ref{eq:entdist}) (ordinate) against \(N\) (abscissa). Squares represent the scenario when the measurement is performed on the first qubit of a chain consisting of nine qubits while circles correspond to the refrigerator where the second qubit is measured. In the former scenario, the nodal site is taken to be the second qubit and in the latter case, it is the first qubit. 
     (b) The position of an unmeasured qubit, \(A_i\), (ordinate) in a spin chain that reaches unit fidelity vs. the position of a qubit on which the measurement is performed, \(A_j^M\) (abscissa). Here \(L=9\). The parameters are same as in Fig. \ref{fig:fig2}. All the axes are dimensionless.  }
     \label{fig:ent_dot_9}
 \end{figure}

{\it Even vs. odd.} When the total number of qubits in the array of the spin chain is even,  repeated measurements on a single qubit project all the subsystems into the corresponding ground state of the individual qubit (in most of even cases). 

This is not the case when $L$ is odd.  Precisely, it is not always possible to cool down all the unmeasured qubits by measuring on an arbitrary qubit, for odd values of $L$, irrespective of the number of rounds, i.e.,  the maximum fidelity in this case saturates to a non-unit fidelity after $N$ rounds of the protocol. The behavior is also independent of the choice of the involved system parameters. Specifically, we find that for odd \(L\), measuring at the qubits which are located in the odd positions, it is indeed possible to cool down all the qubits individually, while if one measures on the qubits belonging to the even site of the spin chain, only unmeasured qubits of even sites attain their respective ground states (see Fig. \ref{fig:ent_dot_9}).  More interestingly, we report  that it is impossible to cool down any qubit of the spin chain if the outcome of the qubit at position $\lfloor \frac{L}{2} \rfloor + 1$ is the ground state of the local Hamiltonian, when $L$ is odd (see  Fig. \ref{fig:ent_dot_9} (ii) when \(A_j^M = 5\)). We will now  argue that the odd-even dichotomy 
can be qualitatively explained from the pattern of entanglement, established between the unmeasured qubits after  the unitary evolution of the entire system and  each round of measurements.

{\it Distribution of entanglement.}  Let the number of sites of the spin chain being $L$ and measurement being performed on $A_{r}^{M}$. We compute the distribution of entanglement \cite{coffman2000, Dhar2017} as
\begin{equation}
    \Sigma_{\mathcal{L}} = \sum_{\substack{i = 1 \\ i \neq r}}^{L} \mathcal{E} (\varrho_{A_{r-1}A_i}),
    \label{eq:entdist}
\end{equation}
where $\mathcal{E}$ is any entanglement measure. In our case, we compute logarithmic negativity \cite{vidal2002, plenio2005} as an entanglement measure and the corresponding quantity is denoted by \(\Sigma_{\mathcal{L}}\). Here $A_{r-1}$ is the nearest neighbor site of the measured qubit which is taken as a nodal site. In a similar argument, \(A_{r+1}\) can also be taken as the nodal party which is the case when the first qubit is measured. 

For odd $L$, when the measurement is performed on the odd qubits, $\Sigma_{\mathcal{L}}$ decreases with increasing number of measurements, thereby demonstrating the action of a refrigerator while $\Sigma_{\mathcal{L}}$ saturates with $N$ when the measurements are performed on the even qubits, implying that not all the qubits are attaining their corresponding ground states. When all the subsystems reach their ground states,  the final state becomes  fully separable and hence, $\Sigma_{\mathcal{L}} = 0$ in the former case while it is not true for the latter case. Fig. \ref{fig:ent_dot_9} (i) clearly mimics this behavior of a chain having nine sites. 

\section{Conclusion}
\label{sec:conclusion}

The idea of miniaturization of devices has revolutionised the field of technology which naturally surpasses classical technology based on classical physics. Therefore, it is a great time to go beyond it and formulate new sketches of old machines on the breadboard of quantum systems. Due to rapid experimental advancements in the quantum domain, one can expect to implement these devices, not only in laboratories but also in an industrial scale.

Summarizing, we proposed a quantum refrigerator model based on multiple measurement schemes. Specifically,  all the subsystems are initially prepared in  thermal states except the one which is to be measured in its ground state of a local magnetic field after a certain interval of time. During dynamics, the system evolves according to an interacting Hamiltonian, a one-dimensional  $XY$ spin chain with long range interaction present between the subsystems  in presence of a transverse magnetic field  and then a measurement is performed at a fixed site. We analyzed the final state of the unmeasured qubits as a consequence of these repeated evolution and measurements. We observed that the projection of a prefixed qubit to its ground state can project the other unmeasured qubits to their minimum energy state, thereby exhibiting the phenomena of cooling. In contrast to previous designing of a quantum refrigerator where a single system is considered as a target system to cool down, in the current proposal, we can cool more than one subsystem by only measuring a single qubit, showing scalability of the cooling mechanism.

Previous understanding of a quantum refrigerator was based on a master equation approach where each qubit or a part of the system is attached to the environment and a target system is taken to be refrigerated. Our approach differs from the previous proposals in the sense that a part of the spin models can be thought of as an environment, and the rest can act as a refrigerator in which  the accessible qubits are used to perform repeated measurements. Moreover, we showed that by wisely choosing system parameters, an efficient quantum refrigerator can be constructed in which less number measurements are required and it can be realized with a nonvanishing success probability. We also found that the position of a qubit in an array of spin-$\frac{1}{2}$ particles chosen for measurement can dictate the amount of cooling, quantified via fidelity between the resulting state and the ground state of the local Hamiltonian and the success probability. We believe that the design of a refrigerator based on measurements may provide a different framework to build quantum thermal machines.

\section*{acknowledgement}
We acknowledge the support from Interdisciplinary Cyber Physical Systems (ICPS) program of the Department of Science and Technology (DST), India, Grant No.: DST/ICPS/QuST/Theme- 1/2019/23. We  acknowledge the use of \href{https://github.com/titaschanda/QIClib}{QIClib} -- a modern C++ library for general purpose quantum information processing and quantum computing (\url{https://titaschanda.github.io/QIClib}) and cluster computing facility at Harish-Chandra Research Institute.

\bibliographystyle{apsrev4-1}
\bibliography{ref}

\end{document}